\documentclass[twocolumn,prb,amsmath,amssymb]{revtex4}
\usepackage{graphicx}
\usepackage{bm}
\begin{document}

\title{Anisotropic magnetic field dependence of 
many-body enhanced\\ electron tunnelling through a quantum dot
}

\author{E.~E.~Vdovin$^{1,2}$, Yu.~N.~Khanin$^{2}$,   
O.~N.~Makarovsky$^{1}$, A.~Patan\`e$^{1}$, L.~Eaves$^{1}$,\\ M.~Henini$^{1}$ C.~J.~Mellor$^{1}$, K.~A.~Benedict$^{1}$ and R.~Airey$^{3}$}

\affiliation{$^{1}$~School of Physics and Astronomy, University of Nottingham, NG7 2RD, UK
\\$^{2}$~Institute of Microelectronics Technology RAS, 142432 Chernogolovka, Russia
\\$^{3}$~University of Sheffield, Department of Electronic \& Electrical Engineering, Mappin Street, Sheffield, S1 3JD}

\begin{abstract} 
We investigate the effect of an applied magnetic field on resonant tunneling of electrons through the bound states of self-assembled InAs quantum dots (QDs) embedded within an (AlGa)As tunnel barrier.  At low temperatures ($\le 2$~K), a magnetic field {\bf B} applied either parallel or perpendicular to the direction of current flow causes a significant enhancement of the tunnel current. For the latter field configuration, we observe a strong angular anisotropy of the enhanced current when {\bf B} is rotated in the plane of the quantum dot layer. We attribute this behavior to the effect of the lowered symmetry of the QD eigenfunctions on the electron-electron interaction.
\end{abstract}

\maketitle

The Fermi-edge singularity (FES) is a many-body interaction effect, which has been observed in a variety of systems, including X-ray absorption in metals \cite{1} and photoluminescence from semiconductor quantum wells (QWs) \cite{2}. Theoretical work by Matveev and Larkin predicted the existence of an FES for the case when electrons tunnel through a localised state in a potential barrier \cite{3}. An electron tunneling into a localized level generates a scattering potential for the electrons in the contact leads. The change of occupation of the localized level during the tunneling leads to a change in the scattering potential and a power-law singularity in the electron tunneling rate. This effect has since been observed in the current-voltage characteristics {\em I(V)} of several semiconductor heterostructure devices \cite{4,5,6,7,8,9}. Experiments on electron tunnelling into isolated donor states embedded in the QW of a double barrier resonant tunnel diode \cite{4} or into self-assembled quantum dots (QDs) in a tunnel diode \cite{5,6,7,8,9} have shown that the FES is strongly enhanced at low temperatures, as predicted by theory \cite{3}. The original Matveev-Larkin model has been extended to conditions which are far from equilibrium \cite{10,11} and to the case of FES in an open planar QD system \cite{12}.

Whilst the temperature and bias dependence of the FES in tunnel diodes are now well understood, experiments involving an additional perturbation on the electron system have revealed interesting features. These include the observation in QD tunneling experiments \cite{5,6,7,8,9} of a strong dependence of the FES on magnetic field, {\bf B}, applied parallel to the direction of current flow, i.e. {\bf B}$\parallel${\bf J}, where {\bf J} is the current density. The tunnel current flowing from a degenerate two-dimensional electron gas (2DEG) through the QD was found to depend in an oscillatory manner on the Landau level filling factor of the 2DEG \cite{5, 6}.  For the case of a three-dimensional (3D) Fermi sea, an enhancement of the current was observed at high B and attributed to the effect of partial spin polarization of the electrons \cite{7, 9}. It has also been shown that a magnetic field reduces the shot noise at the FES \cite{13}.
 
In this paper, we investigate how the tunnel current through a QD at low temperatures is influenced by magnetic field. We examine not only the {\bf B}$\parallel${\bf J} geometry, but also when {\bf B} is perpendicular to {\bf J} For {\bf B}$\perp${\bf J}, we observe an unexpectedly strong angular anisotropy in the FES when {\bf B} is rotated in the plane of the QD layer. Our observations are relevant to recent studies in which the B-dependence of the tunnel current or magnetocapacitance has been used to probe the spatial form of the wavefunctions of electrons confined in the states of a QD \cite{14,15,16} and the effect of electron-electron interactions on the wavefunctions \cite{16,17,18}.

\begin{figure}[t]
\includegraphics[width=3.0 in]{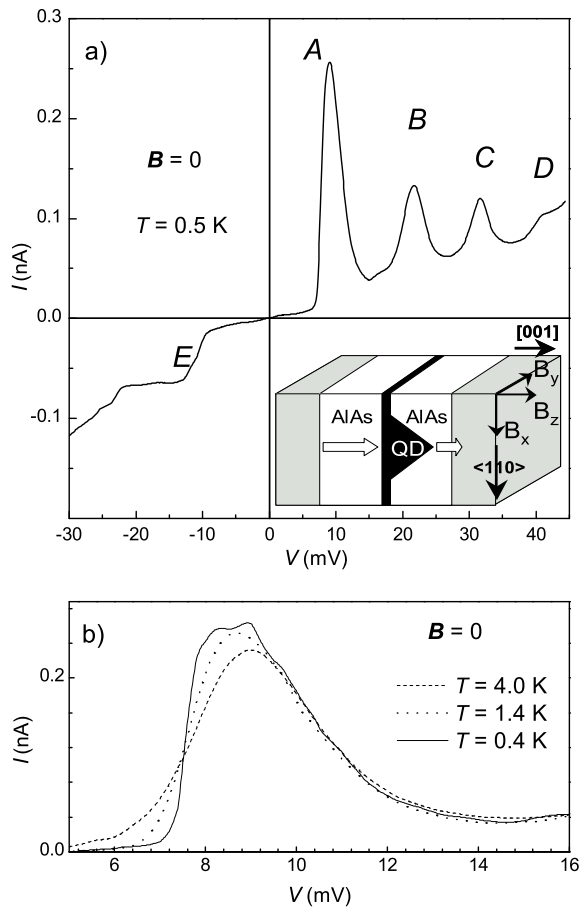}
\caption{(a) {\em I(V)} curve at T = 0.5~K and B = 0~T for a mesa diode of diameter 50~$\mu$m. The inset shows schematically our coordinate scheme relative to the crystallographic axes and electron tunneling through a QD in forward bias. (b) {\em I(V)} curve of the current peak {\em A} at different temperatures and B~=~0~T.}
\end{figure}
 
Our tunnel diodes were grown by molecular beam epitaxy on (001)-oriented Si-doped GaAs substrates with a single layer of self-assembled InAs QDs embedded in the central plane of an Al$_{0.4}$Ga$_{0.6}$As barrier. The detailed layer composition in order of growth is as follows: a Si-doped, 1~$\mu$m GaAs buffer layer (n = $2\times 10^{18}$~cm$^{-3}$); a 40~nm undoped GaAs layer; a 4~nm undoped Al$_{0.4}$Ga$_{0.6}$As barrier layer; a 1.8 monolayer (ML) InAs QD layer; a 4~nm undoped Al$_{0.4}$Ga$_{0.6}$As barrier layer; a 4~nm undoped GaAs layer; a 1.2 ML InAs wetting layer; a 20~nm undoped GaAs layer; a 50~nm GaAs layer (n = $2\times 10^{17}$ cm$^{-3}$); and finally a 0.5~$\mu$m GaAs top cap-layer (n =$2\times 10^{18}$~cm$^{-3}$). The structures were grown at 550$^{\circ}$C except for the InAs layers and the overgrown Al$_{0.4}$Ga$_{0.6}$As and GaAs layers, which were grown at 500$^{\circ}$C. Ohmic contacts were made by deposition and annealing of AuGe/Ni/Au layers. Mesa structures, with diameters of 50~$\mu$m and 200~$\mu$m were fabricated by wet chemical etching. Measurements of the current-voltage characteristics, {\em I(V)}, were made over the temperature range 0.4 to 50~K.  Here we focus on measurements performed in forward bias, for which electrons tunnel from the negatively-biased bottom substrate electrode through the QDs in the barrier and into the positive collector contact, i.e. left to right in Figure~1, inset.

The InAs QD layer creates discrete zero-dimensional electronic states in the tunnel barrier. At zero bias, equilibrium is established by electrons diffusing from the doped GaAs layers towards the QDs. The partial charging of the QDs produces depletion layers in the GaAs layers adjacent to the Al$_{0.4}$Ga$_{0.6}$As barrier \cite{14}. An applied voltage, {\em V}, shifts the QD energy levels with respect to the Fermi energy, $E_F$, of the GaAs emitter layer. When a particular QD level coincides with $E_F$, resonant tunneling of electrons leads to a sharp increase in current.
 
Figure~1(a) shows {\em I(V)} curves for a 50~$\mu$m diameter mesa measured at T = 0.5~K. Sharp peaks labeled {\em A, B, C} and {\em D}, are observed in forward bias between 8 and 50~mV. These resonances can be observed up to T~=~50~K. This peak structure is sample-specific but, for a given sample, it is exactly reproducible even after thermal cycling of the sample following an extended period (months) at room temperature. We attribute each peak to resonant tunneling of electrons from the emitter Fermi sea into a discrete state of a QD. The difference between the forward and reverse bias {\em I(V)} curves arises from the asymmetry of the device structure: the barrier and InAs wetting layer grown on the top side of the QDs has a higher transmission coefficient than that of the substrate side \cite{19}. Thus, in forward bias, the time-averaged occupancy of the QD state remains small since the tunneling rate from the emitter to the QD is much smaller than the tunneling rate out of the QD into the collector. In reverse bias, the step-like increase of the current occurs because of Coulomb charging of the QDs \cite{20}.
  
In the following, we will concentrate on resonant peak labeled {\em A} as the FES in a magnetic field both the parallel and the perpendicular to the current is observed on the threshold of this resonance. The form of resonance {\em A} in Figure~1(b) is approximately triangular with a sharp onset at the low bias edge, $V_0$ = 7.2~mV, and a tail extending to higher bias. This is consistent with energy-conserving electron tunneling from a degenerate 3D Fermi gas in the GaAs emitter into a 0D state in the barrier \cite{21}. When the bias is increased beyond 12~mV, resonant tunneling ceases as the QD state drops below the conduction band edge in the emitter. The low bias-edge of peak {\em A} broadens with increasing T from 0.4 to 4.2~K due to thermal broadening around the chemical potential of the emitter Fermi sea. Fitting the temperature dependence of the low bias edge gives an electrostatic leverage factor $f = 0.44\pm 0.05$, which is the fraction of total applied bias dropped between the emitter and the QD layer in the barrier. Let's note, that a singularity was not observed in absence of the magnetic field even at the lowest temperature (see Fig.1(b)). This is in contrast with the case for electron tunneling through shallow donors\cite{4} where a pronounced FES is observed at B = 0. Insignificant increasing of the amplitude of the resonance {\em A} we attributed to decreasing of the thermal broadening of the Fermi distribution function in the emitter. However, the fine structure of the current peak is observed with decreasing temperature down to 0.4~K. This fine structure is caused by fluctuations of the local density of states in the emitter.

\begin{figure}[t]
\includegraphics[width=3.0 in]{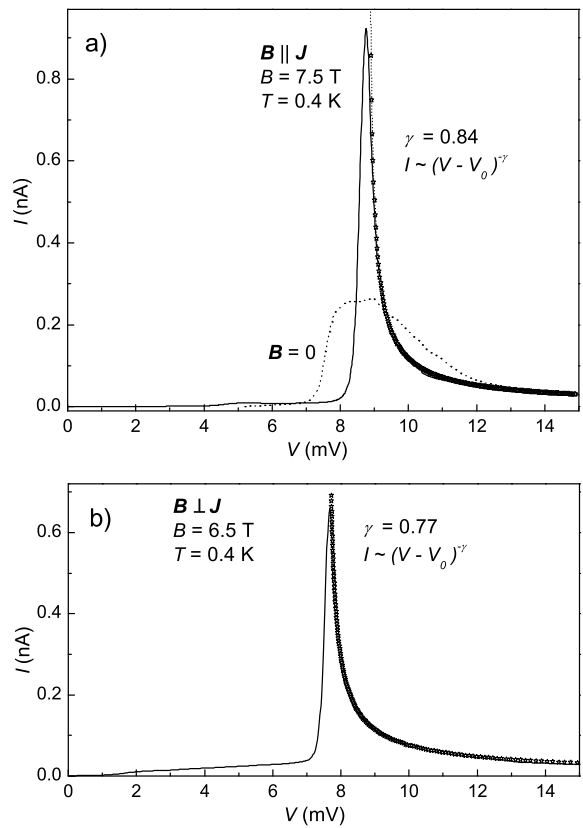}
\caption{(a) {\em I(V)} curve for {\bf B}$\parallel${\bf J} at 7.5~T at  T = 0.4~K. The stars represent the curve $I\sim (V - V_0)^{-\gamma}$ with $\gamma\sim 0.84$. The dashed line is the {\em I(V)} curve at  T = 0.4~K and B = 0~T. (b) {\em I(V)} characteristic for {\bf B}$\perp${\bf J} at 6.5~T at T = 0.4~K. The stars represent the curve $I\sim (V - V_0)^{-\gamma}$ with $\gamma = 0.77$. 
}
\end{figure}
  
At T = 0.4~K, the lineshape and intensity of peak {\em A} is strongly field-dependent, see Figure~2(a) and (b). Decreasing of the temperature to T = 0.4~K leads to sharp increasing of the front of the resonant feature at 8~mV (peak {\em A}) near the threshold of tunneling when lowest QD state is resonant with Fermi energy of emitter. Full width half maximum (FWHM) values of the singularity near the peak {\em A} as small as 0.3~mV has been measured, corresponding to 130~$\mu$eV in energy scale. The shape of this sharp current peak can be described by a steep ascent and a more moderate decrease of the current towards higher voltages. We related the unexpected rapid enhancement of the low voltage resonance in magnetic field at low temperature with manifestation of the interaction-induced Fermi-edge singularity in the tunneling current via a localized state. In order to analyze this many-body current we investigated the slope of the peak. For peak {\em A}, the {\em I(V)} curve with {\bf B}$\parallel${\bf J} at 7.5~T, see Figure~2(a), has the characteristic form of an FES-enhanced tunnelling resonance with a sharp, well-defined low bias edge at $V_0$ = 8.5~mV followed by a tail-off of the current at higher bias.  The tail can be fitted accurately by a Matveev-Larkin type power law, $I\sim (V - V_0)^{-\gamma}$ with $\gamma = 0.84 \pm 0.05$. This value is in qualitative agreement with that obtained recently by Hapke-Wurst et al. \cite{7} in a similar QD structure for the same field orientation. We observe the same type of behaviour at 0.4~K with {\bf B}$\perp${\bf J}, i.e. {\bf B} in the (001) crystallographic plane, see Figure~2(b); here a best fit to the high bias tail is obtained with $\gamma = 0.77 \pm 0.05$.  These values of $\gamma$ are considerably larger than that ($\gamma = 0.22$) derived from experiments involving electron tunneling through a shallow donor impurity\cite{4} at  B = 0.  We also note that they are beyond the range of the standard FES model, which is strictly applicable only for $\gamma\ll 1$ (See Ref.\cite{3}).

\begin{figure}[b]
\includegraphics[width=3.1 in]{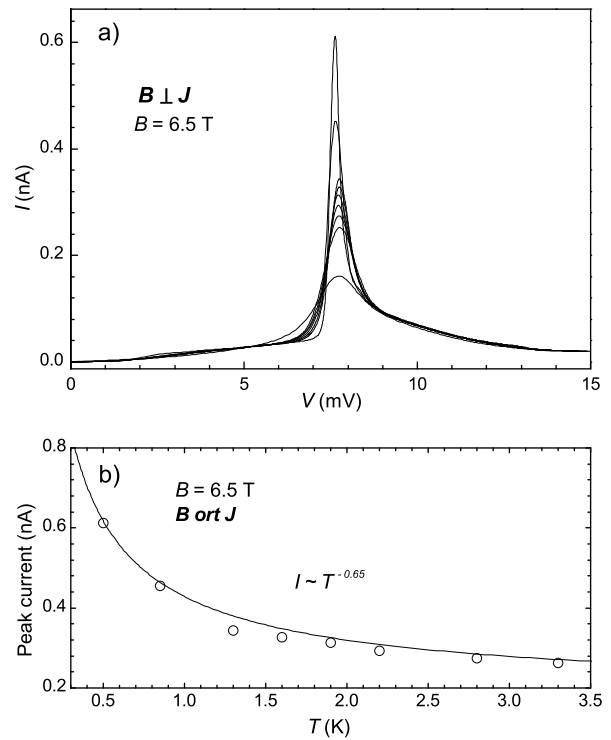}
\caption{(a) Temperature dependence of peak {\em A} in {\em I(V)} with {\bf B}$\perp${\bf J} at 6.5~T. (b) The T-dependence of the maximum of the peak current (stars) and the curve $I \sim T^{-\gamma}$  with $\gamma = 0.65$.}
\end{figure}

A different way to determination of the origin of the anomalous enhancement of the tunneling through quantum dot is a temperature-dependent experiment. As predicted in Ref.\cite{26}, the fact that the area under {\em I-V} curve around the resonance increases with decreasing temperature strongly indicates that the temperature dependence in this case is of a many-body nature. A similar temperature behavior is demonstrated by our experimental curves clearly (Fig.3(a)). Moreover temperature dependent measurements offer an additional way to determine the edge exponent \cite{7,27}. As shown in Fig.3(b) the amplitude of the peak {\em A} decreases according to a power law $I\sim T^{-\gamma}$  with $\gamma\sim 0.65$, a value in good agreement with that derived from the bias dependence of current at constant T. Thus the analyze of the temperature dependences independently shows that the anomalous enhancement of the tunneling current has been attributed to a many-body contribution which arises due to the strong interaction of a tunnelling electron with the Fermi sea in the emitter.

The extreme values of the edge exponent of $\gamma\sim 0.7$ can't been explain by theory of the FES valid for $\gamma\ll 1$  without additional modifications \cite{3}. A similar on observed by us dramatic enhancement of the FES has been reported in Ref.\cite{7} in very high magnetic fields (up to 28~T) parallel to the current. In this paper was proposed the adapted model of FES and have shown that the interaction between a localized charge on the QD and the electrons in the Landau quantized emitter leads to FES if only the lowest Landau level in the 3D emitter is occupied due to dramatic Fermi phase shifts. This results in edge exponents $\gamma\sim 0.5$ in high magnetic fields when the electrons in the lowest Landau level of the emitter is mainly spin polarized. In our experiment the values of the Fermi energy in the 3D emitter at the bias near the peak {\em A} is 2.4~meV and it is approximately in 6 times less when in Ref.\cite{7}. As a consequence electrons in the emitter at 8~T is totally spin polarized and we can observed FES at relatively slow magnetic field.

\begin{figure}[t]
\includegraphics[width=3.3 in]{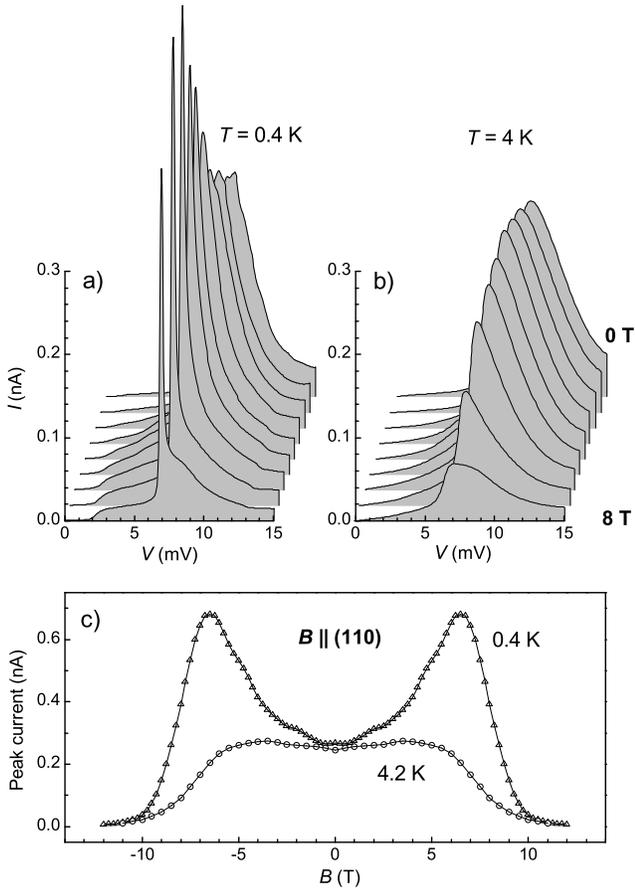}
\caption{Magnetic field dependence from 0 to 8~T of peak {\em A} in {\em I(V)} with {\bf B}$\perp${\bf J} at: (a) T = 0.4~K and (b) T = 4.2~K. {\bf B} is oriented along one of the $\left\langle 110\right\rangle$ crystalline axes in the (001) plane. (c) Dependence of peak current on magnetic field B applied along $\left\langle 110\right\rangle$ at T = 0.4 and 4.2~K.}
\end{figure}

In order to investigate further the nature of the FES for the {\bf B}$\perp${\bf J} configuration, we have carried out a series of angular anisotropy measurements for different orientations of {\bf B} relative to the crystalline axes in the plane of the tunnel barriers. We now consider the magnetoanisotropy effects observed for {\bf B}$\perp${\bf J}.  Figure~4 compares the B-dependence of peak {\em A} for two different temperatures, 0.4~K(a) and 4~K(b).  The field is applied parallel to one of the $\left\langle 110\right\rangle$ axes in the (001) plane (to within $\pm 5^{\circ}$). As shown in Figure~4(c), a strong increase of the peak current is observed at T = 0.4~K with increasing {\bf B} up to 7~T. When {\bf B} is increased further, the peak current decreases and is quenched above 10~T. In contrast, at T = 4~K the magnetic field leads to an approximately monotonic decrease of the amplitude of the peak current, similar to the behavior observed previously for other QD devices investigated at 4~K \cite{14, 15}.

The data at T = 4~K can be understood in terms of the action of the magnetic field on a tunneling electron.  The Lorentz force imparts to the electron an in-plane momentum given by $\hbar k = -edB \times${\^z} , where {\em d} is the effective tunneling distance along the normal ({\em{\^z}}) to the barrier plane. In the absence of interaction effects, the tunnel current into a particular QD state is then proportional to $|{\varphi}_{QD}(k)|^2$, where ${\varphi}_{QD}(k)$ is the single particle eigenfunction of the state in {\em k}-space \cite{15}. Thus, by carrying out a series of measurements of the tunnel current as a function of {\bf B} for a range of orientations in the (001) plane, the probability density in Fourier space of the ground and excited states of a QD can be mapped \cite{14,15,16,22}.

\begin{figure}[b]
\includegraphics[width=3.4 in]{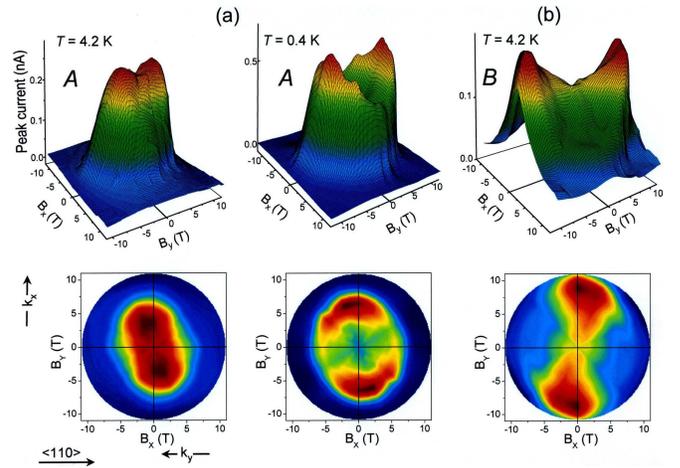}
\caption{Color plots of the anisotropy for different orientations of {\bf B} in the (001) crystalline plane of the peak current for: (a) resonance {\em A} at T = 4.2~K and 0.4~K; (b) resonance {\em B}. The $B_x$ and $B_y$ axes are parallel to the two orthogonal $\left\langle 110\right\rangle$ axes of the plane.  B = 10~T corresponds to an in-plane {\em k}-vector equal to 0.5~nm$^{-1}$.}
\end{figure}
  
The magneto-anisotropy of peak {\em A} at 4~K when {\bf B} is rotated in the (001) plane is plotted in Figure~5(a). This indicates that the QD state associated with this resonance has a bell-shaped probability density with a broad maximum centered at B = 0 and k = 0. From the layer composition of the device, we deduce that $d = 30\pm 5$~nm (See Ref.\cite{8}).  Therefore, an in-plane field of 10~T imparts to the tunneling electron an in-plane k-vector equal to 0.5~nm$^{-1}$. The overall form of the magnetoanisotropy plot of peak {\em A} is characteristic of a ground state QD eigenfunction of ``1{\em s}-like'' character. However, there is some weak, yet sharp, structure close to B = 0 and a pronounced elongation of the probability density in momentum space with major and minor axes close to the two $\left\langle 110\right\rangle$ directions of the (001) plane.
 
At 0.4~K we observe a similar two-fold magneto-anisotropy of the FES enhancement of peak {\em A}, see Figure~5(a) and Figure~6. The striking feature is that the enhancement occurs predominantly when {\bf B} is applied along one of the $\left\langle 110\right\rangle$ directions of the (001) plane. The contour plot no longer resembles the bell-shaped probability density of a 1{\em s}-like state in momentum space, but it contains a significant component with 2{\em p}-like character. This type of symmetry is also revealed in the anisotropy plots for the higher bias peaks {\em B} and {\em C}. The plot for peak B has the form of the probability density of a 2$p_x$-like excited state of a QD in which the {\em p}-lobes are oriented along one of the $\left\langle 110\right\rangle$ directions of the (001) plane with maxima at $k_x \approx \pm 0.3$~nm$^{-1}$. We cannot state definitively that peaks {\em B} and {\em C} are the excited states of the dot that gives rise to peak {\em A}. However, the dots in the ensemble should have the same symmetry. In addition, the probabilities densities of peaks {\em A, B} and {\em C} are fully consistent with those expected for the orthogonal ground and 2{\em p} excited states of QDs with an elongated confinement potential.

\begin{figure}
\includegraphics[width=3.0 in]{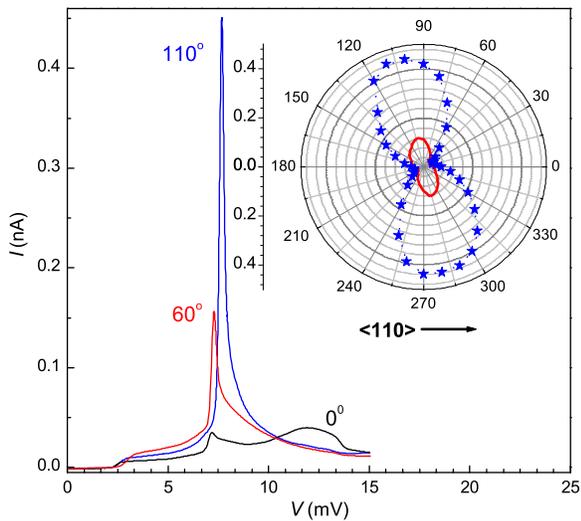}
\caption{{\em I(V)} characteristic at the perpendicular to the current magnetic field of 8~T at T = 0.4~K. Inset: angular dependence of the peak current (stars -- at T = 0.4~K, solid line -- at T = 4~K).}
\end{figure}

The angular measurements have allowed us to receive two completely unexpected results, which one could not be forecast within the framework of the existing theories. At first, as it is visible from a Fig.6, the amplitude FES has very strong angular dependence. Secondly, the anisotropy of FES is reflection of an anisotropy of wave function of the quantum dot. Should be noted that that FES and wave function of the QD demonstrate identical type of the anisotropy, despite of strong difference of dependences of the peak current from value of magnetic field in high and low temperature cases (as it's shown in Fig.4(c)).

The FES arises from the response of the electrons in the emitter Fermi sea to the change in charge on the dot when an electron tunnels into and out of it. The anisotropy of the FES enhancement indicates that this many-body interaction is strongly influenced by the anisotropic confinement potential of the QDs. In particular, the appearance of a partial 2$p_x$-like character in the lower temperature (T = 0.4~K) anisotropy plot of peak {\em A} in Figure~4(a) may arise from a virtual tunneling process through the first-excited state (2$p_x$-like) and then down to the ground state (1{\em s}-like) of the dot, mediated by interactions with electrons in the emitter Fermi sea.
 
The effect of many-body interactions on electron tunneling through QDs has recently been considered by Rontani \& Molinari both for the case of magnetotunnelling and scanning tunneling microscopy (STM) experiments \cite{17,18}.  Their calculations indicate that the quasi-particle wavefunctions of electrons measured by these techniques can differ significantly from those of the single particle states of the QD, i.e. the tunnel current can be modified by correlation effects within the dot when it is already occupied. Some evidence of this type of effect has been reported in magnetocapacitance studies \cite{16}. In our experiments, the temperature- and voltage-dependences of the peak A measured below 4~K have all the characteristics of an FES-induced involving interactions with electrons in the emitter. This indicates that, at sufficiently low temperature, interactions between the charge state of the dot and the electrons near the Fermi energy in the contact lead can be comparable to those between electrons confined within the dot itself.  Similar FES effects could also be observed in STM imaging experiments of QD eigenstates, although the relatively high temperatures (4~K) STM data reported in the recent literature \cite{23,24,25} may have hindered this observation. It is also worth noting that in STM imaging experiments of QDs, the barrier thickness between the tip (contact lead) and the dot is smaller than the size of the dot and much smaller than the tunneling length ({\em d} = 30~nm) in our experiments. These smaller tunneling distances could enhance many-body interaction effects with the Fermi sea of the STM tip.

In conclusion, we observe a strong enhancement of the low temperature ($\le 2$~K) tunnel current through the ground state of an InAs QD in the presence of a magnetic field. An unexpectedly strong angular anisotropy in the FES is observed when the magnetic field is rotated in the plane of the QD layer and attributed to the effect of the lowered symmetry of the QD eigenfunction on the electron-electron interaction. These results are relevant to recent studies of the effect of many-body interactions on electron tunneling through discrete quantum states. 
 
\begin{acknowledgments}
The work is partly supported by EPSRC (UK), INTAS grant RFBR 06-02-16556, the Royal Society and the SANDiE Network of Excellence of the European Commission (NMP4-CT-2004-500101). The authors thank A.V.~Khaetski and I.V.~Larkin for useful discussions, and V.V.~Belov for technical assistance.
\end{acknowledgments}


\begin{thebibliography}{text}
\bibitem{19} Our photoluminescence measurements indicate that in forward bias the InAs wetting layer (WL) on the collector side of the barrier is unoccupied by electrons. The proximity of this WL to the QD layer allows efficient electron tunneling from the dots through the WL and then into the collector contact.
\end{thebibliography}
\end{document}